\documentclass[nopubldata]{lpor2014}
\usepackage{blindtext}
\usepackage{hyperref}
\hypersetup{%
  colorlinks,
  plainpages=false,
  pdfpagelabels,
  breaklinks,
  pdfview=Fit,
  bookmarksopen,
  bookmarksnumbered,
  linkcolor=black,
  anchorcolor=black,
  citecolor=black,
  filecolor=black,
  urlcolor=LPRred
  }%
\graphicspath{{./figs/}}
\setpages[]{}
\setvolume[0]{0}
\setyear{2011}%
\setdoi{201100000}%
\oddheadlogotext{}{}
\usepackage{soul}
\usepackage{color}
\usepackage{float}
\usepackage{fancybox}

\newcommand{\lsq}{\left[}
\newcommand{\rsq}{\right]}

\usepackage{pifont}% http://ctan.org/pkg/pifont
\newcommand{\cmark}{\ding{51}}%
\newcommand{\xmark}{\ding{55}}%
\usepackage{CJKutf8}
\usepackage{textcomp}
\usepackage{multicol}
\usepackage{stfloats}
\usepackage{physics}
\makeatletter
\newcommand*{\addFileDependency}[1]{% argument=file name and extension
  \typeout{(#1)}
  \@addtofilelist{#1}
  \IfFileExists{#1}{}{\typeout{No file #1.}}
}
\makeatother

\usepackage{hyperref}
\usepackage{xr-hyper} 

\begin{document}

%\titlefigure{}
\abstract{%
High energy single- to few-cycle terahertz pulses enable the exploration of the frontier of science, such as electron acceleration,  strong-field physics, and spectroscopy. One important method of generating such terahertz pulses is to use the tilted-pulse-front (TPF) technique. However, due to the non-collinear phase-matching, the large angular dispersion leads to a spatial and temporal break-up of the optical pump, limiting the terahertz generation efficiency and reducing the few-cycle character of the generated terahertz fields. To reduce the effects caused by angular dispersion, multiple schemes with discrete pulse-front-tilt have been suggested. We propose a terahertz generation scheme, where a spatio-temporally chirped (STC)  pump pulse is utilized. With our 2D+1 numerical model, we perform a systematic comparative study of the conventional TPF scheme, three discrete TPF schemes, and the STC scheme. This model predicts smaller optimal interaction length and significantly smaller conversion efficiency compared to simple 1D models. We conclude that the STC scheme delivers spatially homogeneous few-cycle terahertz pulses with the highest conversion efficiency. Additionally, given a short interaction length, the discrete TPF schemes cannot outperform the continuous ones. In general, this work gives guidance to choose the most appropriate setup for a given terahertz experiment.}

\title{Tilted-pulse-front schemes for terahertz generation}

\author{Lu Wang\inst{1,2,*}, Gy\"orgy T\'oth \inst{3},J\'anos Hebling \inst{3,4,5} and Franz K\"artner\inst{1,2,6}}%
\authorrunning{L.Wang}
\mail{\email{lu.wang@desy.de}}

\institute{%
Center for Free Electron Laser Science (CFEL), Deutsches Elektronen-Synchrotron, Hamburg, 22607, Germany
\and
Universit\"{a}t Hamburg, Department of Physics, Hamburg,  22761, Germany
\and
University of P\'ecs, Institute of Physics, 7624 P\'ecs, Hungary
\and
MTA-PTE High-Field Terahertz Research Group, 7624 P\'ecs, Hungary
\and
University of P\'ecs, Szent\'agothai Research Centre, 7624 P\'ecs, Hungary
\and
The Hamburg Centre for Ultrafast Imaging (CUI), Hamburg, 22761, Germany
}

\maketitle

\section{Introduction}

The last several decades have witnessed a tremendous increase in the laser-based, high-energy terahertz applications \cite{salen2019matter} such as spectroscopy \cite{davies2002development}, strong field terahertz physics \cite{kampfrath2011coherent,schubert2014sub}, particle acceleration \cite{zhang2018segmented}, electron spin manipulation \cite{kampfrath2011coherent} and phonon resonance studies \cite{bakker1992observation}. Optical rectification is an important method for terahertz generation, where an ultrashort pump laser pulse induces a strong dipole moment via the second-order nonlinear effect. During the interaction of the terahertz and the optical beams, a repeated energy down conversion of pump photons (cascading effect) is possible, leading to broadband terahertz pulse emission with an efficiency close to, or even above, the Manley-Rowe limit \cite{hemmer2018cascaded}. The "tilted-pulse-front" (TPF) technique, a phase matching (PM) method for terahertz generation by optical rectification, or more generally for a non-collinear interaction, brings new possibilities to generate high energy terahertz pulses. In this technique, the intensity front of the optical pump (OP) is tilted with respect to the phase front \cite{hebling1996derivation}. The generated terahertz propagates perpendicularly to the TPF \cite{hebling2002velocity}. Due to the non-collinear phase-matching, frequency downshifted optical components generated via the cascading effect, possess large angular spread. This leads to a spatial and temporal break-up of the optical pump, limiting the terahertz generation efficiency and reducing the few-cycle character of the generated terahertz fields. Furthermore, due to the material absorption at terahertz frequencies, limited damage threshold of the nonlinear material and the low terahertz photon to pump photon energy ratio, generating high energy terahertz pulses is challenging.

{Multiple schemes have been suggested to generate pulse-front-tilt. In this article, we focus on two types of pulse-front-tilt: "continuous", where the TPF forms a continuous plane and "discrete", where the TPF is achieved by discrete beamlet structure.} The conventional grating (CG) scheme was proposed and demonstrated in 2002 by J. Hebling et.al.\cite{hebling2002velocity}, where a diffraction grating induces angular dispersion onto the optical pump pulse, leading to a pulse front tilt. Shortly after, in 2004, the pulse front tilt caused by spatio-temporal chirp (STC) was proposed by S. Akturk et.al. \cite{akturk2004pulse}. In this case, the pulse front tilt is generated by propagating a transversely linear-chirped optical pulse through a dispersive medium. In this method, no angular dispersion occurs. However, this method has not been used to generate terahertz pulses. In 2016, BK. Ofori-Okai et.al. \cite{ofori2016thz} demonstrated a setup consisting of a stair-step reflective echelon (RES) structure instead of a reflective optical grating. The echelon produces a discretely tilted pulse front, decreasing the negative effect of large angular dispersion \cite{ofori2016thz}. In 2017, a multistep phase mask (MSPD) scheme was proposed by Y. Avetisyan et.al. \cite{siliconmask}. This scheme splits a single input beam into many smaller time-delayed "beamlets". Compared with the grating method, it decreases the negative effects of the angular dispersion and eliminates the necessity of the imaging optics. In the same year, L. P\'alfalvi et.al. performed numerical studies of a nonlinear echelon (NLES) slab \cite{palfalvi2017numerical}, where a stair-step echelon-faced nonlinear crystal is used instead of a nonlinear prism. It is expected that this scheme produces good-quality, symmetric terahertz beams. Recently, the corresponding experiment was demonstrated by PS. Nugraha et.al. in 2019 \cite{nugraha2019demonstration}. The CG and STC schemes generate contineous TPF whereas the RES, MSPM and NLES schemes generate discrete TPF.
 \begin{figure*}[bt]
\centering
{\includegraphics[width=0.7\textwidth]{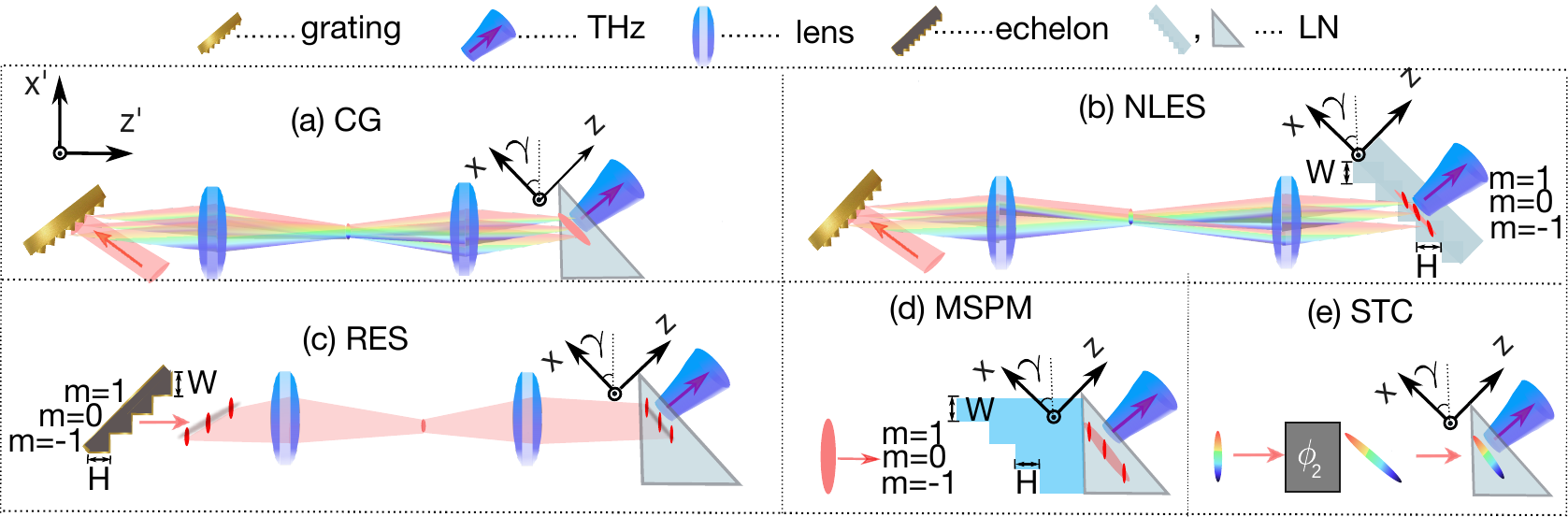}} 
\caption{Schematic drawings of different tilted-pulse-front schemes. (a-e) are the configurations for conventional grating (CG), nonlinear echelon (NLES), reflective echelon (RES), multistep phase mask (MSPD) and spatio-temporal chirp (STC) schemes respectively. (b-d) represent the discrete TPF schemes whereas (a) and (e) represent the continuous TPF schemes. The step numbers of the structures are labeled by $m$. The apex of the LN crystal is located at $x=0$. The z=0 coordinate is defined at the location where the center of the OP starts to interact with the LN crystal.\label{setup_pic}}
\end{figure*}

In this article, a 2-D+1 (x,z,t) numerical model is used to investigate the effectiveness of the various tilted pulse front schemes. The effect of the vertical dimension ($y$ dimension in Fig. \ref{setup_pic}) is of minor importance in this study \cite{lu3d}. The nonlinear material lithium niobate (LN) is chosen due to its large second order optical nonlinearity. The conversion efficiency and the spatial distribution of the generated terahertz electric fields are presented for the aforementioned 5 schemes. With thorough discussions of the advantages and disadvantages of each scheme, this work gives guidance to choose the most appropriate setup for a given terahertz experiment.

\section{The investigated setups}
The illustrations for different tilted-pulse-front schemes are presented in Fig. \ref{setup_pic}. For a fair comparison, the OP peak fluence, the OP beam size $\sigma_x'$ and the size of the beamlets \emph{at the LN input surface} (i.e after the optical elements and the imaging system) are set the same for all the schemes. The OP propagates along the $z'$ direction and the terahertz propagates along the $z$ direction. The analytical forms of the OP electric fields inside the LN crystal and the second order nonlinear polarisation can be found in the supporting information.  

In schemes (a-c) in Fig. \ref{setup_pic}, the imaging systems are chosen such that the image of the grating (echelon) overlaps with the tilted pulse front inside the LN crystal \cite{tokodi2017optimization}. This is considered to be the optimal imaging condition and the analytical expressions can be found in Eqs. (1-3) of the supporting information. The first lens has a fixed focal length $f_1=300$ mm and the second one, close to the LN, has a focal length $f_2$. {In scheme (e), the TPF is achieved by applying second-order dispersion $\phi_2$ to a spatially chirped OP. Consequently, no initial angular dispersion is induced.} The simulation parameters are listed in Table \ref{parameter}.
\vspace{-0.3cm}
\begin{table}[h] 
\centering {\caption{ Simulation parameters \label{parameter}}}
\medskip
\begin{tabular}{c|c}
\hline
parameter name & value\\
\hline
wavelength $\lambda_0$ & 1018 nm\\
 beam size $\sigma_x'$  & 0.5 mm $\&$ 4 mm\\
 %\hline
  grating period $d$ & 1/1500 mm \\
%\hline
focal length $f_1$ &  300 mm \\
%\hline
  phase matching angle $\gamma$ & $64.8^{\circ}$ \\
%\hline 
phase matching frequency $\Omega_0$ & 2$\pi \times$ 0.3 THz \\ 
%\hline
pulse duration (FWHM) $\tau_0$ & 500 fs \cite{fulop2012generation} \\
% \hline
temperature  &300 K\\
%\hline
  peak fluence& $\sqrt{10 \tau_{0} (\text{fs}) }$\, mJ/cm$^2$  \cite{ravi2016pulse}\\
%\hline
 THz absorption  $(300\, \text{K},0.3 \,\text{THz})$ & 7/cm \cite{unferdorben2015measurement} \\
 %\hline
 NLES scheme W, H & 97 \textmu m , 206 \textmu m  \\
 %\hline
 RES scheme W, H & 150 \textmu m \cite{ofori2016thz}, 229 \textmu m  \\
 %\hline
 MSPM scheme W, H & 97 \textmu m , 1000 \textmu m  \\ 
 %\hline
 STC scheme $\phi_2$ & 0.045 $\text{ps}^2$   \\
  \hline
 \end{tabular}
\end{table}
\vspace{-0.2cm}
\section{Results}
\subsection{Efficiency}
\begin{figure}[h]
\centering
{\includegraphics[width=0.42\textwidth]{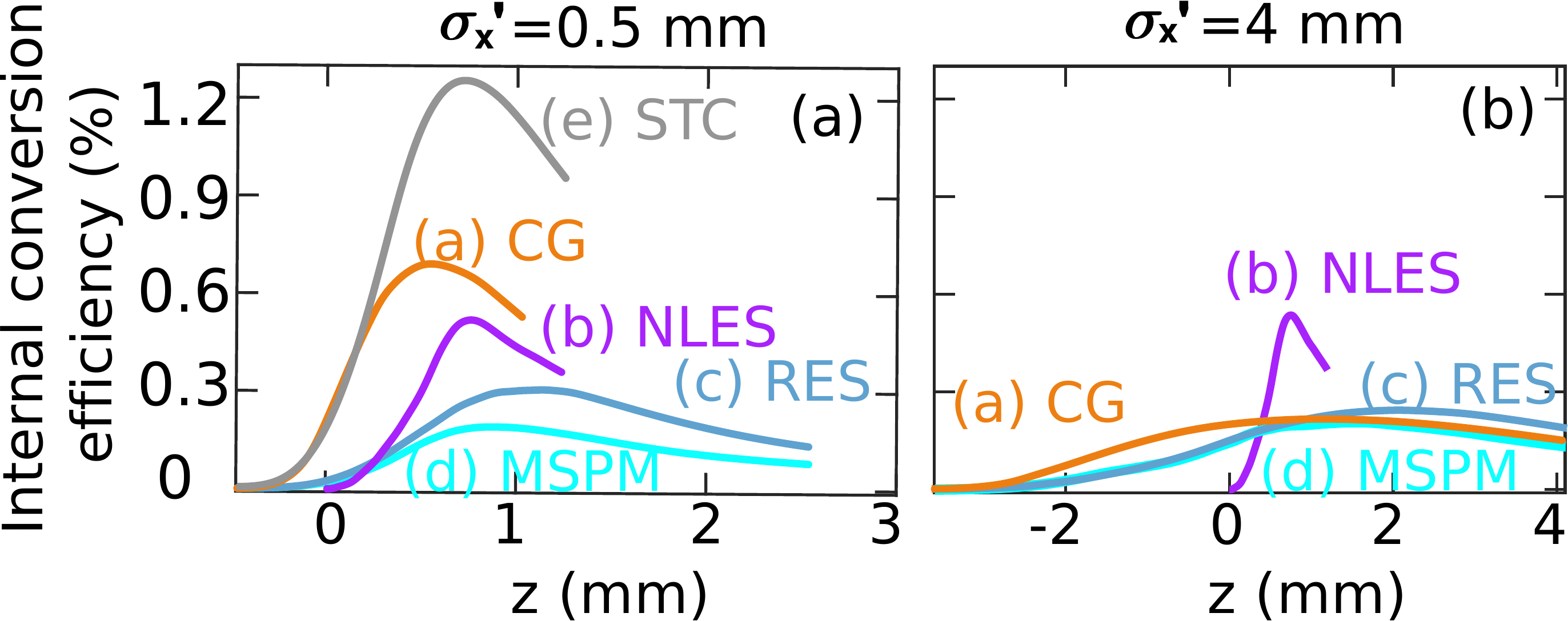}} 
\caption{The internal conversion efficiency (without out-coupling loss) of different tilted-pulse-front schemes versus the interaction length ($z$). (a) and (b) correspond to 0.5 mm and 4 mm OP beam sizes respectively. The $z=0$ coordinate is defined at the location where the center of the OP starts to interact with the LN crystal. \label{eff}}
\end{figure}

Two OP beam sizes, 0.5 mm and 4 mm, corresponding to short and long interaction lengths are analyzed. The two beam sizes are chosen to elucidate the impact of upfront angular dispersion in connection with the short and longer interaction lengths, by solving the 2D+1 coupled nonlinear wave equations \cite{lu3d}. It can be seen in Fig. \ref{eff} that the CG scheme loses its advantage for longer interaction length, due to the maximum initial angular dispersion among all the 5 scheme. For the NLES scheme with high input OP fluence, the optimal effective length is $L_\text{eff} \approx 1/\alpha$. When $\sigma_x'> \sin{(\gamma)} L_\text{eff}$, the efficiency and the optimal interaction length is nearly independent of the OP beam size. Since the NLES slab creates beamlets at the LN input surface, all of the beamlets experience an almost identical condition i.e similar walk-off distance. Thus, the generated terahertz pulse properties are not related to the OP beam size, which is the unique character of the NLES scheme compared to all the other schemes.

\begin{figure*}[b!]
\centering
\includegraphics[width=1\textwidth]{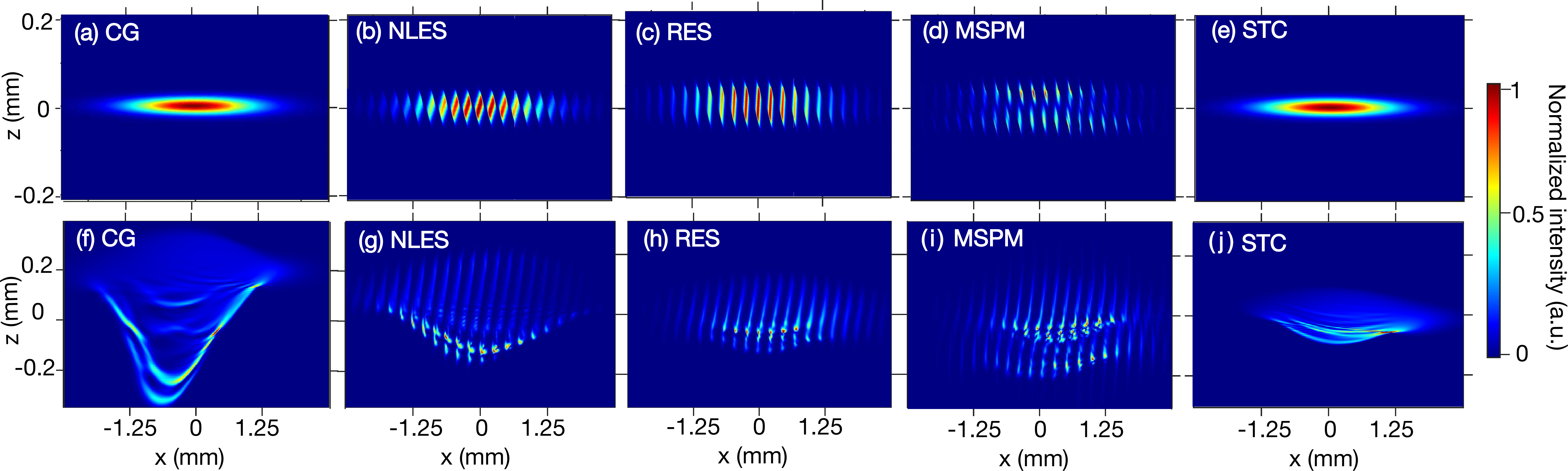} 
\caption{The OP ($\sigma_x'=0.5$ mm) intensity distributions versus the transverse dimension x for different TPF schemes are shown. (a-e) show the corresponding input OP intensity distributions at the LN input surface. (f-j) show the corresponding OP intensity distributions after a given propagation distance ($1.2$ mm). The intensity distributions are shown in x-z coordinates where the distribution along z dimension is linearly related to the distribution in time and $\sigma_x=\sigma_x'/\cos{(\gamma)}$. The center of the OP is chosen to be $x=0,\, z=0$, for convince of representation.  \label{op_tx}}
\end{figure*}

 The RES and the MSPM schemes are similar from the point of view that the PM is achieved entirely by generating time-delayed beamlets. However, despite the similar interaction length, RES outperforms MSPM in terms of efficiency due to the imaging system and the absence of dispersion in the mask material. In the MSPM scheme, the OP experiences diffraction and dispersion which modifies the spatial and temporal profiles of the beamlets (see Fig. \ref{op_tx}(d)). The STC scheme delivers the highest conversion efficiency, since the OP contains zero initial angular dispersion and continuous TPF. However, the STC scheme is only applicable for a small OP beam size due to the limitation in OP bandwidth (see section \ref{each_scheme}, scheme (e) for more detail).

One can see that for the discrete TPF schemes (NLES, RES, MSPM), the conversion efficiency does not strongly dependent on the OP beam size (or interaction length). With a short interaction length, the conversion efficiency of the discrete TPF schemes can not exceed the continuous TPF schemes. Since the beamlet structure leads to a discrete PM condition i.e the entire arrangement (envelope) of the beamlets forms the TPF, while each individual beamlet itself has an offset with respect to the perfect pulse-front-tilt surface (see supporting information). Additionally, among all the 3 discrete TPF schemes, the NLES scheme delivers the highest conversion efficiency since the beamlets are more tilted towards the pulse-front-tilt surface.

Figure \mbox{\ref{op_tx}} shows the impact of angular dispersion onto the OP. The initial angular dispersion condition of the 5 schemes is CG$>$ NLES $>$RES, MSPM $>$STC. Figure \mbox{\ref{op_tx}}(f-j) suggest that with a given propagation distance, the OP with larger initial angular dispersion suffers more from temporal and spatial pulse break-up. This spatial and temporal break-up limits the terahertz generation efficiency and reduces the few-cycle character of the generated terahertz fields (see Fig. \mbox{\ref{simu}} for details).

\subsection{Terahertz out-coupling loss}
Due to the large terahertz refractive index inside LN ($n(\Omega_0)= 5.2$), a significant energy loss of the terahertz pulse propagating from the LN into the air is inevitable. The maximum Fresnel transmission T at the desired terahertz frequency is only T=$1-(n(\Omega_0)-1)^2/(n(\Omega_0)+1)^2 = 0.54 $. However, because of the distortion of the terahertz beams, T is much smaller and is very sensitive to the spatial distribution of the terahertz beam, which is largely dependent on the fluence and size of the OP. The T at the peak of each efficiency curve in Fig. \mbox{\ref{eff}} are calculated and the corresponding results are listed in Table \ref{fresnel}.

\begin{figure*}[b]
\centering
\includegraphics[width=1\textwidth]{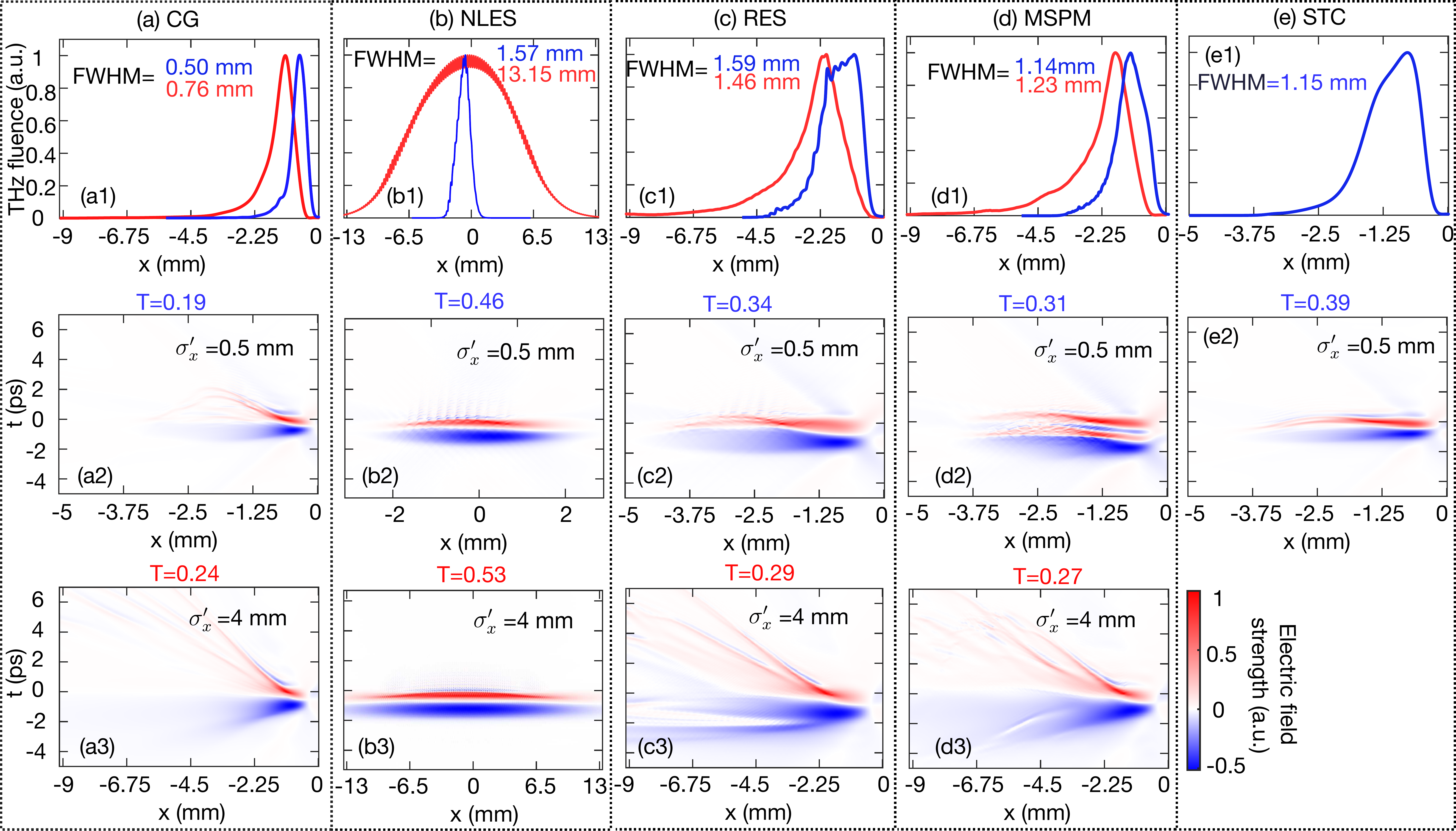} 
\caption{Numerical results of the terahertz electric fields generated by different schemes at the peak of the efficiency curves in Fig. \ref{eff} are shown. The LN apex is located at $x=0$. The figure labels a-e correspond to schemes (a)-(e), respectively. (a1-e1) represent the output terahertz fluence with OP beam sizes 0.5 mm (blue) and 4 mm (red) respectively. The terahertz electric field distributions versus $x$ generated by 0.5 mm and 4 mm OP are shown in (a2-e2) and (a3-d3) respectively. The terahertz energy transmission at the exit surface of the LN is denoted by T.   \label{simu}}
\end{figure*}

For all the mentioned schemes apart from NLES, the terahertz fields, generated by OP with lower fluence or smaller beam size, express less spatial inhomogeneity and thus have higher transmission. Besides, larger OP beam size leads to lower terahertz frequency due to longer interaction length, since the absorption of the LN increases with the terahertz frequency (see supporting information). In principle, larger beam size would reduce the angular divergence and thus enhances the transmission. However,  increasing the OP beam size does not lead to an strong increase of the terahertz beam size (see Fig. \mbox{\ref{simu}}). Additionally, the resulting larger terahertz beam has worse beam quality due to the variation of interaction length along the transverse dimension $x'$ caused by the prism geometry. On the other hand, one can see from Table \mbox{\ref{fresnel}} that the NLES scheme has the highest transmission (especially for large OP beams) owing to the spatially homogeneous terahertz beam quality.

\begin{table}[h]
\centering
\caption{Terahertz energy transmission (T) and the external conversion efficiency ($\eta_e$) at the LN output surface. The highest $\eta_e$ for a given OP beam size is marked by a black box. \label{fresnel}}
\medskip
\begin{tabular}{c| c c | c c}
 \hline
 &   \multicolumn{2}{c|}{$\sigma_x'=$0.5 mm} & \multicolumn{2}{c}{$\sigma_x'=$4 mm} \\
\hline 
 & T& $\eta_e$ ($\%$) &T& $\eta_e$ ($\%$)\\
\hline
 (a) CG & 0.19& 0.12 & 0.24& 0.05 \\
 %\hline
 (b) NLES & 0.46 & 0.24 & 0.53 & \Ovalbox{0.29}\\
 %\hline
 (c)RES& 0.34 & 0.11&0.29&0.07\\
 %\hline
 (d) MSPM  &0.31&0.06  & 0.27& 0.06\\
 % \hline
 (e) STC& 0.39 & \Ovalbox{0.48} & - & -\\
\hline
\end{tabular}
\end{table}

\section{Analyses of each scheme}\label{each_scheme}
To further understand the terahertz generation process with the aforementioned schemes, the terahertz electric fields at the peak of the efficiency curves in Fig. \ref{eff}, generated by 0.5 mm and 4 mm OP beam sizes are presented in the following sections. Within the $x'-z'$ plane, two phase matching conditions need to be satisfied, in order to generate terahertz pulses efficiently (see derivation of PM conditions in the supporting information). One is shown in Eq.\mbox{(\ref{vm})}, where $n_g$ is the group velocity refractive index of the LN at center frequency $\omega_0$ and $\gamma$ is the TPF angle. This condition ensures that the projection of the pump velocity equals to the terahertz velocity \mbox{\cite{hebling2002velocity}}. This is also known as velocity matching, which applies to all the schemes discussed.  
\begin{equation}
   n(\Omega_0)\cos{(\gamma)}=n_g, \label{vm} 
\end{equation}
The other PM condition varies according to the scheme and is given in the following sections respectively. This condition is responsible for creating tilted (average) pump intensity front with the appropriate tilt angle $\gamma$.  

\subsection*{Scheme (a): conventional grating (CG).}
 The ease of this setup, the high pulse energies and the controllability of the terahertz properties has made this scheme a strong candidate for generating high energy terahertz pulses approaching the milijoule range \cite{fulop2014efficient}. However, the transversal asymmetry of the interaction, in combination with the cascading effect, result in terahertz beams with non-uniform spatial distributions. The CG scheme utilizes the grating induced angular dispersion to form a pulse front tilt \cite{hebling1996derivation}. The angular dispersion has two effects on the OP away from the imaging plane i.e increase the OP pulse duration and decrease the TPF angle \cite{martinez1986pulse}. These two effects are more pronounced for larger angular dispersion (or broadband OP), leading to the minimum interaction length in the CG scheme compared with all the other schemes.  The PM condition is given by Eq. (\mbox{\ref{cg_pm}}), where $c$ is the speed of light, $\beta=2\pi/\lsq \omega_0 \cos{(\theta_o)}d \rsq$ is the first order angular dispersion induced by the grating and $\theta_o$ is the grating output angle.
\begin{eqnarray}
  \tan{(\gamma)}=\beta c f_1/(f_2n_g) \label{cg_pm}
\end{eqnarray}
 Fig. \ref{simu}(a1-a3) indicate that, compared to the 0.5 mm OP, the terahertz field generated by 4 mm OP suffers from more transversal spread in terms of fluence and temporal distribution. Furthermore, most of the terahertz energy is contained in the single-cycle region of the terahertz electric field. By increasing the OP spot size by $\times 8$, the terahertz beam size increases only by $50\%$. This is the main reason of the efficiency drop for large OP beam size as shown in Fig. \ref{eff}.   

\subsection*{Scheme (b): nonlinear echelon (NLES).}
This method is beneficial for generating high energy, large size and spatially homogeneous terahertz beams. The step size of the LN nonlinear echelon along the $x'$ and $z'$ dimensions are represented by W and H respectively (see Fig. \ref{setup_pic}(b)). The PM condition is given by
\begin{eqnarray}
 \lsq n_g\tan{(\gamma)}-{c\beta f_1/}{f_2}\rsq \text{W}=\text{H}(n_g-1)  \label{nles_pm}
\end{eqnarray}

 In order to ensure that the terahertz pulse propagates perpendicular to the entrance and exit surfaces of the plan-parallel LN slab, the condition $\text{H}/\text{W}=\tan{(\gamma)}$ need to be satisfied \cite{palfalvi2017numerical}. With this condition, Eq. (\ref{nles_pm}) reduces to $\tan{(\gamma)=c\beta f_1/f_2}$. Compared with the CG scheme in Eq. (\ref{cg_pm}), the NLES scheme requires less angular dispersion. Additionally, with different input OP pulse duration, the optimal effective length may differ. Thus, the thickness of the NLES slab should vary accordingly. However, for a given NLES with fixed thickness, this can be adjusted by adapting the pump fluence.

It can be seen from Fig. \ref{simu}(b1-b3) that by changing the OP beam size the generated terahertz beam size changes accordingly. Besides, as long as $\sigma_x'> \sin{(\gamma)} L_\text{eff}$ is satisfied, the terahertz generation efficiency and the terahertz spectra are independent of the OP beam size, which is a unique property of this scheme. This property enables the possibility of generating high energy terahertz pulses by simply increasing the pump energy and the OP beam size. Additionally, the electric fields show strong single-cycle character homogeneously along the entire transverse dimension $x$.

\subsection*{Scheme (c): reflective echelon (RES).}
The echelon step sizes in $z'$ and $x'$ dimensions are represented by H and W respectively (see Fig. \ref{setup_pic}(c)). The temporal delay between the two neighboring beamlets is $2\text{H}/c$. This remain unchanged after the imaging system, whereas the transversal spacing between adjacent beamlets  reduces from W to $f_2/f_1$W. The PM condition given by Eq. (\ref{res_pm}).
\begin{eqnarray}
\tan{(\gamma)}=2\text{H}f_1/(\text{W}n_gf_2) \label{res_pm}
\end{eqnarray}
It can be seen from Fig. \ref{simu}(c1-c3) that the electric fields distributions are very similar to the CG scheme. However, the terahertz beam size is far larger ($\sim \times 2$) than the one produced by the CG.

\subsection*{Scheme (d): multi-step phase mask (MSPM).}
In this scheme, no imaging system is required. The delay of each beamlet is generated by different propagation lengths inside the mask stripes. In the simulation, silica with refractive index of $n=1.45$ \cite{malitson1965interspecimen} is chosen as the mask material due to its low dispersion at the OP wavelength. The phase mask step size in $x'$ and $z'$ dimensions are W and H respectively (see Fig. \ref{setup_pic}(d)). This scheme is very similar to the RES scheme. However, the imaging system in the RES scheme guarantees that at the image plane, each beamlet experiences the same condition. In this scheme, the diffraction modifies the beamlets envelope drastically, leading to poor terahertz electric field quality. It can be seen in the comparisons of Figs. \ref{simu}(c1-c3) and \ref{simu}(d1-d3) that an imaging system is necessary. This scheme can not outperform the RES scheme. The PM condition is given in Eq. (\ref{mspm_pm}). 
\begin{eqnarray}
\tan{(\gamma)}=(n-1)\text{H}/(\text{W}n_g) \label{mspm_pm}
\end{eqnarray}

\subsection*{Scheme (e): spatio-temporal chirp (STC).}
{In this scheme, no initial angular dispersion is present, leading to a maximum conversion efficiency among all 5 schemes discussed. However, the input OP pulse has to be broadband and a ELI-ALPS SYLOS laser can be an ideal option. 

The full width half maximum (FWHM) of the overall input OP spectral bandwidth is presented in Eq. (\mbox{\ref{fwhm}}), which corresponds to a $\sim$10 fs transform limited pulse. For a fair comparison, the local transform limited pulse duration ($\tau_0$) at a given $x'$ position is chosen to be 350 fs. Thus, the local pulse duration remains 500 fs after the chirp. }
% \begin{figure}[H]
% \centering
% \includegraphics[width=0.27\textwidth]{pictures/st_input.pdf} 
% \caption{The input OP spectral distribution versus the transverse coordinate $x'$ is shown. \label{simu_st_input}}
% \end{figure}
\begin{eqnarray}
   f_{\text{FWHM}}=\sigma_x'\sqrt{{2\log{(2)}}({v^2\tau^2}/{2}+{1}/{\sigma_x'^2})}/{\tau\pi} \label{fwhm}
\end{eqnarray}
In Eq. (\ref{fwhm}), $v$ is the spatial chirp rate and $\tau=\tau_0/\sqrt{2\log{2}}$. The PM condition is presented in Eq. (\ref{stc_pm}). The definition of OP as a function of $\phi_2$ and $v$ can be found in Eq. 19 of the supporting information.   
\begin{eqnarray}
\tan{(\gamma)} =v\phi_2c/n_g\label{stc_pm}
\end{eqnarray}
One can also choose a smaller input OP bandwidth together with a larger $\phi_2$. However, this leads to a temporal broadening of the OP at each spatial point, which is not in favor of the terahertz generation process. 

With the given OP beam size ($\sigma_x'=$0.5 mm), this scheme delivers the highest conversion efficiency and a spatially homogeneous few-cycle terahertz field. However, the disadvantage is that the input OP must contain a large bandwidth. Additionally, since the bandwidth of the OP is related to the input beam size (see Eq. (\ref{fwhm})), this scheme is not applicable to large OP beam size.

\section{Conclusion}
Due to the non-collinear phase-matching, frequency downshifted optical components generated via the second-order effect, possess large angular spread. This leads to a spatial and temporal break-up of the optical pump, limiting the terahertz generation efficiency and reducing the few-cycle character of the generated terahertz fields.  Additionally, large angular diffraction of the terahertz reduces the out-coupling efficiency of the terahertz fields which reduces the overall efficiency further. The simulations suggest that with lower OP input intensity, the terahertz electric field is closer to the single-cycle format along the $x$ dimension and the terahertz beam size increases (see supporting information). 

The CG and STC schemes form continuous TPF, where the PM condition is fulfilled along the entire transverse dimension. For the schemes related to beamlets (NLES, RES and MSPM), the entire beamlet-train forms the TPF with the required tilt angle. However, each individual beamlet itself has an offset with respect to the perfect TPF surface. Given a short interaction length, the discrete TPF schemes can not outperform the continuous TPF schemes in terms of efficiency. The MSPM scheme delivers the lowest efficiency and for large OP beam size, the efficiency is comparable to the RES scheme. We do not recommend MSPM scheme for small OP beam size. Among all 3 discrete TPF schemes, the NLES has the best performance in terms of efficiency and terahertz beam quality.
 
Schemes CG, NLES and RES, are applicable for a large range of parameters such as OP energy, bandwidth and beam size. Within these three schemes, the CG favors smaller interaction length (small beam  size) and narrower OP bandwidth due to the large angular dispersion. NLES has a potential of delivering large and homogeneous terahertz beam because of the plan-parallel shape of the LN crystal and the smaller imaging errors in comparison to the CG scheme. Additionally, the generated terahertz spectrum does not depend on the OP beam size. The NLES and RES schemes require manufacturing \textmu m sized structures, which is time consuming and prone to manufacturing errors.

Due to zero initial angular dispersion, the STC scheme delivers the highest conversion efficiency and spatially homogeneous few-cycle terahertz field. However, the OP pulse has to be broadband. Due to the spatial chirp, the bandwidth scales linearly with the OP beam size, making this scheme not applicable to large OP beam sizes.

\begin{table}[H]
\centering
\caption{Comparison of different schemes. The symbols \cmark \cmark, \cmark and \xmark \,represent recommend, neutral and not recommend respectively.  \label{experiment}}
\medskip
\begin{tabular}{c| c c c c c}
 \hline
 & (a)  & (b)  & (c) & (d) & (e) \\
 & CG & NLES & RES & STC&  MSPM\\
\hline
 THz quality & \xmark & \cmark \cmark  & \cmark &\cmark \cmark& \xmark\\
 %\hline
 $\sigma_x'$ scalability & \cmark & \cmark \cmark  & \cmark &\xmark& \cmark\\
 %\hline
parameter flexibility & \cmark \cmark & \cmark  & \cmark \cmark &\cmark & \xmark\\
%\hline 
efficiency &  \cmark & \cmark \cmark  & \cmark & \cmark \cmark & \xmark\\
\hline
\end{tabular}
\end{table}
A summary of the advantages and disadvantages of each scheme is listed in Table \ref{experiment}. Please note that the item "$\sigma_x'$ scalability" is equivalent to large input OP energy since the maximum fluence is limited by the damage threshold of the LN crystal.

\section*{Supporting Information}
 Supporting Information is available from the Wiley Online Library or from the author.\\

\noindent The code is available among reasonable request. To access the code, please contact (no space):\\ lu (dot) wangphysics (at) gmail (dot) com

\section*{Acknowledgments}
Lu Wang would like to thank Dr. Dongfang Zhang for his inspiring discussions and patience, IMPRS for the support in both scientific research and in life and the rainy days in Hamburg for giving her no choice but to stay indoor and concentrate on research. Gy\"orgy T\'oth would like to thank the support of the J\'anos Bolyai Research Scholarship of the Hungarian Academy of Science. This project has received fundings from European Union's Seventh Framework Program (FP7/2007-2013) through the Synergy Grant AXSIS (609920) and the Hamburg Cluster of Excellence 'CUI: Advanced Imaging of Matter' of the Deutsche Forschungsgemeinschaft (DFG) - EXC 2056 - project ID 390715994.

\section*{Conflict of Interest}
The authors declare no conflict of interest.

\section*{Keywords}
terahertz generation, tilted-pulse-front, numerical modeling, phase-matching
\bibliographystyle{lpr}
\bibliography{sample}

\end{document}

%\titlefigure{}
\abstract{%
This  document  provides  supporting information  to "Tilted-pulse-front schemes for terahertz generation". The optimal imaging conditions, the analytical forms of the optical pump (OP) electric fields and the derivation of the phase-matching (PM) conditions are given. The spectra and fluence of the terahertz and the OP corresponding to the main text are presented. }

\title{ Tilted-pulse-front schemes for terahertz generation: Supporting Information}

\author{Lu Wang\inst{1,2,*}, Gy\"orgy T\'oth \inst{3},J\'anos Hebling \inst{3,4,5} and Franz K\"artner\inst{1,2,6}}%
\authorrunning{L.Wang}
\mail{\email{lu.wang@desy.de}}

\institute{%
Center for Free Electron Laser Science (CFEL), Deutsches Elektronen-Synchrotron, Hamburg, 22607, Germany
\and
Universit\"{a}t Hamburg, Department of Physics, Hamburg,  22761, Germany
\and
University of P\'ecs, Institute of Physics, 7624 P\'ecs, Hungary
\and
MTA-PTE High-Field Terahertz Research Group, 7624 P\'ecs, Hungary
\and
University of P\'ecs, Szent\'agothai Research Centre, 7624 P\'ecs, Hungary
\and
The Hamburg Centre for Ultrafast Imaging (CUI), Hamburg, 22761, Germany
}

\maketitle

\section{Introduction}
 Simulations of terahertz generation with the CG scheme have been developed by M. I. Bakunov et al. \cite{bakunov2008terahertz,bakunov2011terahertz} via a 2D+1 numerical model. However, this model doesn't include the back conversion of the terahertz to the optical pump (OP). Consequently, the effective length and conversion efficiency are overestimated. Later on, the interaction between the optical pump and the terahertz pulse was included in the 2D+1 model along with a one lens imaging system by K. Ravi et al. \cite{ravi2015theory}. The one-lens imaging system has larger imaging errors and induces more terahertz divergence \cite{tokodi2017optimization}, compared with the telescope imaging system. In the work presented in the main text, a 2-D+1 (x,z,t) numerical model is used to investigate the tilted pulse front schemes since the effect of the $y$ dimension is negligible \cite{lu3d}. The numerical tool is based on the fast Fourier transform beam propagation method (FFT-BPM) \cite{feit1978light} and the split-step Fourier method. The combination of these two methods reduces computational cost compared to the finite difference time-domain (FDTD) method which is very accurate but computationally demanding. The nonlinear polarization terms are calculated in the frequency domain. The entire electric field is updated by the 4th order Runge-Kutta in the propagation direction of the terahertz radiation ($z$). The analytical results are shown with the assumption that each beamlet is of Gaussian shape. In the numerical calculation, the beamlets maintain the shape of the overall Gaussian envelope.  The out-coupling loss is calculated by Fourier transforming the terahertz filed into the reciprocal domain ($k_x$). The transmission according to the Fresnel law is calculated for every single frequency.

\section{Imaging system}
 The optimal imaging conditions for different schemes are the following, where $\theta_o$ is the grating output angle of OP and $n_g$ is the optical group refractive index at the center wavelength $\lambda_0$ in lithium niobate (LN), $n_0$ is the optical refractive index at $\lambda_0$ in LN, $\gamma$ is the tilted-pulse-front (TPF) angle (see Fig. 1 in the main text), $d$ is the grating period. The first lens has a focal length $f_1$ and the second one, close to the LN, has a focal length $f_2$.
 \vspace{0.2cm}
\begin{align}
\hline
&\text{conventional grating (CG), scheme (a):  }\label{image1} \\
\hline
\nonumber \\
&\sin{(\theta_o)} = \frac{1}{2}\lsq \frac{-n_0\lambda_0}{\tan^2(\gamma)n_gd} +\sqrt{\left(\frac{n_0\lambda_0}{\tan^2(\gamma)n_gd}\right)^2+4}\,\, \rsq , \nonumber \\
 &\frac{f_2}{f_1}=\frac{\lambda_0}{d\cos(\theta_o)\tan(\gamma)n_g}=0.613  \nonumber \\
 \nonumber \\
 \hline
& \text{nonlinear echelon (NLES), scheme (b): } \label{image2}\\
\hline
\nonumber \\
&\sin{(\theta_o)} = \frac{1}{2}\left[\frac{-n_0\lambda_0}{\tan^2(\gamma)(n_0-1)d}+\sqrt{\left( \displaystyle \frac{n_0\lambda_0}{\tan^2(\gamma)(n_0-1)d}\right)^2+4}\, \,\right], \nonumber \\
 &\frac{f_2}{f_1}= \frac{\lambda_0}{d\cos(\theta_o)\tan(\gamma)}=1.057 \nonumber \\
 \nonumber \\
 \hline 
&\text{reflective echelon (RES), scheme (c):} \label{image3} \\
\hline
\nonumber \\
& \sin{(\theta_o)} =\sqrt{\frac{1}{1+2n_0/\tan^2(\gamma)n_g}} , \,\, \frac{f_2}{f_1}= \sqrt{\frac{2}{n_0\,n_g}}=0.646 \nonumber \\
\nonumber \\
\hline \nonumber   
\end{align}

\section{Analytical and numerical results}
In this section, the analytic expressions for the electric field behind the input surface of the LN and the corresponding second order polarization are given. Apart from the spatio-temporal chirp (STC) scheme, the input OP at the OP propagation frame can be defined as Eq. (\ref{e_input}), where $\sigma'$ is the input OP beam size before the optical system and $\Delta \omega=\omega-\omega_0$. The parameter $\sigma_x'$ in Table. 1 in the main text is the OP beam size at the LN input surface i.e after optical elements and the imaging system. For the convenience of representation, $x'=0$ is chosen to be the center of the OP beam. 
\begin{eqnarray}
E(\omega,x',z')=&&E_0\exp{\ls \frac{-\Delta \omega^2\tau^2}{4} \rs}\exp{\ls \frac{-x'^2}{2\sigma'^2} \rs} \nonumber \\
&&\times \exp{\ls -i\omega z'/c \rs}.\label{e_input}
\end{eqnarray}

The ansatz for the terahertz electric field is shown in Eq. (\ref{thz_e_sup}), where $n_t$ is the terahertz refractive index in LN. The wave equation for terahertz generation can be written in the form of Eq. (\ref{wave_e}) where the corresponding second order polarization is defined in Eq. (\ref{pol}).
\begin{eqnarray}
E(\Omega,x',z')=A(\Omega,x',z')\exp{\lsq -i\Omega n_t\sin{(\gamma)}x'/c \rsq} &&\nonumber \\
 \times \exp{\lsq -i\Omega n_t\cos{(\gamma)}z'/c \rsq} && \label{thz_e_sup} \\
 \nonumber \\
 {\partial^2A(\Omega,x',z') }/{\partial x'^2}+{\partial^2 A(\Omega,x',z') }/{\partial z'^2}-\frac{2i\Omega n_t}{c} &&\nonumber \\
\times \large{[} \cos{(\gamma)}{\partial A(\Omega,x',z')}/{\partial z'}+\sin{(\gamma)}{\partial A(\Omega,x',z') }/{\partial x'}\large ] && \nonumber \\
=-\frac{\Omega^2}{c^2} P(\Omega,x',z')+\frac{i\Omega n_t}{c} A(\Omega,x',z')\alpha(\Omega)&& \label{wave_e}\\
 \nonumber \\
P(\Omega,x',z')=\chi^{(2)}\int_0^{\infty}E(\omega+\Omega,x',z')E^*(\omega,x',z') d \omega &&    \nonumber  \\
\times \exp{\lsq i\Omega n_t\sin{(\gamma)}x'/c\rsq} \exp{\lsq i\Omega n_t \cos{(\gamma)}z' /c\rsq} && \label{pol}
\end{eqnarray}

\subsection*{(a) conventional grating (CG)}
The electric field behind the LN input surface is shown in Eq. (\ref{sup_cg1}) where, $\beta = 2\pi/ [\omega_0 \cos{ (\theta_o)}d]$ is the first order angular dispersion induced by the grating, $\theta_i$ is the grating incident angle of the OP. Accordingly, the second order polarization for terahertz generation is shown in Eq. (\ref{sup_cg2}). Since $\sigma_x'$ is the OP beam size at the LN input surface, from Eq. (\ref{sup_cg1}), it can be seen that $\sigma_x'=\sigma'f_2 \cos{(\theta_o)}/f_1\cos{(\theta_i)}$.
\begin{eqnarray}
  E(x',\omega)=E_0\sqrt{\frac{f_1\cos{(\theta_i)}}{f_2\cos{(\theta_o)}}}\exp{\ls \frac{-\Delta \omega^2\tau^2}{4} \rs}&& \nonumber \\
  \times \exp{\ls \frac{-x'^2f_1^2\cos{(\theta_i)^2}}{2\sigma'^2f_2^2\cos{(\theta_o)^2}} \rs} \exp{\ls-i\beta \Delta \omega x' f_1/f_2 \rs}&& \nonumber \\
  \times \exp \ls {-i\omega n_\text{op}(\omega)z'}/{c} \rs && \label{sup_cg1}
\end{eqnarray}
\begin{eqnarray}
   P^{(2)}(\Omega)=\chi^{(2)}\frac{E_0^2\sqrt{2\pi}f_1\cos{(\theta_i)}}{\tau f_2\cos{(\theta_o)}} \exp{\ls \frac{-\Omega^2\tau^2}{8} \rs}&&\nonumber\\ 
   \times  \exp \lsq  \frac{-x'^2 f_1^2\cos{(\theta_i)^2}}{\sigma'^2 f_2^2 \cos{(\theta_o)^2} } \rsq  \exp{\lsq iz'\Omega (n_t\cos{(\gamma)}-n_g)/c \rsq} &&\nonumber \\
\times \exp{\lc i\Omega x' \lsq n_t \sin{(\gamma)}/c-\beta f_1/f_2 \rsq \rc} \hspace{0.2cm}  \label{sup_cg2} &&
\end{eqnarray}
From Eq. (\ref{sup_cg2}), it can be seen that in order to achieve the maximum polarisation, the last two exponential terms need to be 1, which leads to the following two {PM} conditions.
\begin{eqnarray*}
 n_t\cos{(\gamma)}=n_g \nonumber  &&\\
\tan{(\gamma)}=\beta c f_1/(f_2n_g)   &&
\end{eqnarray*}
The terahertz spectra generated by different OP beam sizes are presented in Fig. \ref{simu_cg_spec}. One can see that, the terahertz pulse generated by larger OP beam size possesses lower center frequency due to the absorption through the longer interaction length. Figure \ref{simu_cg_eff} shows the dependence of the conversion efficiency, the terahertz fluence and the spectra on OP fluence. Lower OP fluence leads to larger terahertz beam size. However, the terahertz spectrum doesn't vary much with respect to fluence. In this scheme, the angular dispersion is the dominating limiting factor for effective interaction length. By reducing the energy, the effective length doesn't vary much.{ The OP output spectrum and fluence correspond to Figs. 2(a) and 4(a2) are given in Fig. \mbox{\ref{cg_op}}}.  
\begin{figure}[h]
\centering
\includegraphics[width=0.5\textwidth]{cg_spec.pdf} 
\caption{(a) represents the terahertz spectra generated by 0.5mm (blue) and 4 mm OP (red) respectively. The spatial dependence of the terahertz spectral densities are shown in (b) and (c) respectively.\label{simu_cg_spec}}
\end{figure}
\vspace{-0.5cm}
\begin{figure}[h]
\centering
\includegraphics[width=0.5\textwidth]{cg_energy_compare.pdf} 
\caption{With a given OP beam size ($\sigma_x'=$0.5 mm), (a) represents the internal terahertz conversion efficiency with pump fluence 70.7 mJ/cm$^2$, 35.4 mJ/cm$^2$ and 17.7 mJ/cm$^2$ respectively. The terahertz fluences and corresponding spectra at the peak of each efficiency curve are shown in (b) and (c). The out-coupling energy transmission is denoted by T. \label{simu_cg_eff}}
\end{figure}
\vspace{-0.5cm}
\begin{figure}[h]
\centering
\includegraphics[width=0.5\textwidth]{cg_op.pdf} 
\caption{The OP spectral density (a), spectrum (b) and the fluence (c) are presented. \label{cg_op}}
\end{figure}

\subsection*{(b) nonlinear echelon (NLES)}
The step size of the LN nonlinear echelon along $x'$ and $z'$ dimensions are represented by W and H respectively (see Fig. 1(b) in the main text). By setting  $\sigma_m' \sim \text{ W}/2$, $q_0=i\pi\sigma_m'^2/\lambda_0$ and $q_m=q_0-m\text{H}+m\text{H}/n_0$, the electric field behind the LN input surface is given in Eq. (\ref{sup_nles1}) where $\sigma_x'=\sigma'f_2 \cos{(\theta_o)}/f_1\cos{(\theta_i)}$. The corresponding second order polarization can be found in Eq. {({\ref{sup_nles2}})}. 
\begin{eqnarray}
E(\omega,x',z')=E_0\sum _{m=-\infty} ^{\infty}\sqrt{\frac{q_0f_1\cos{(\theta_i)}}{q_m f_2\cos{(\theta_o)}}}\exp{\ls \frac{-\Delta \omega^2\tau^2}{4} \rs}  && \nonumber \\
\times \exp{\ls \frac{-x'^2f_1^2\cos{(\theta_i)^2}}{2\sigma'^2f_2^2\cos{(\theta_o)^2}} \rs}  \exp \lsq \frac{-ik_0(x'-m\text{W})^2}{2q_m}  \rsq  &&    \nonumber \\
 \times \exp{\ls-i\beta \Delta \omega x' f_1/f_2 \rs}\exp \ls \frac{-i\omega n_\text{op}(\omega)z'}{c} \rs  &&  \nonumber\\
\times \exp \lsq -i\omega m\text{H}\ls n_\text{op}(\omega)-1\rs/c \rsq   \hspace{0.5cm} \label{sup_nles1} &&
\end{eqnarray}

\begin{eqnarray}
P(\Omega,x',z')=\frac{E_0^2\chi^{(2)}\sqrt{2\pi}f_1\cos{(\theta_i)} }{\tau f_2\cos{(\theta_o)} }\sum _{m=-\infty} ^{\infty}\abs{\frac{q_0}{q_m }} && \nonumber \\
\times \exp{\ls \frac{- \Omega^2\tau^2}{8} \rs} \exp{\ls \frac{-x'^2f_1^2\cos{(\theta_i)^2}}{\sigma'^2f_2^2\cos{(\theta_o)^2}} \rs} &&  \nonumber \\
 \times \exp \lsq \frac{-k_0^2\sigma_m^2(x'-m\text{W})^2}{\abs{q_m}^2}  \rsq  \exp \lsq  i\Omega z'(n_t\cos{(\gamma)}-n_g)/c \rsq  && \nonumber \\
 \times \exp  \{ i\Omega [ n_t\sin{(\gamma)}x'-c\beta x' f_1/f_2 - m\text{H} ( n_g-1) ] /c \} \hspace{0.5cm} \label{sup_nles2} &&   
\end{eqnarray}
The two PM conditions are listed below.
\begin{eqnarray}
 && n_t\cos{(\gamma)}=n_g  \nonumber \\
 &&\lsq n_g\tan{(\gamma)}-{c\beta f_1}/{f_2}\rsq x'=m\text{H}(n_g-1) \label{sup_nles_pm}
\end{eqnarray}
The PM conditions require that the last two exponential terms in Eq. \ref{sup_nles2} equal to 1. {Since the $m$ can only be integers (-1, 0 1 ...), the last term of Eq. (\mbox{\ref{sup_nles_pm}}) can only be satisfied for discrete $x'$ values. The entire arrangement of the beamlets forms the TPF, while each beamlet itself is not completely aligned to the TPF plane. In order to ensure that the terahertz pulse propagates perpendicular to the entrance and exit surfaces of the plan-parallel LN slab, the condition H$/$W $= \tan {(\gamma) }$ need to be satisfied. This condition together with $m=1$, $x'=$W, Eq. (\mbox{\ref{sup_nles_pm}}) reduces to the PM condition given in the main text $\lsq n_g\tan{(\gamma)}-{c\beta f_1/}{f_2}\rsq =\text{H}(n_g-1)/\text{W}$ }. Figure \ref{simu_nles_spec} shows the terahertz spectra generated by different OP beam size. In this scheme, the terahertz spectral shape is independent of the OP beam size. 

\begin{figure}[H]
\centering
\includegraphics[width=0.5\textwidth]{nles_spec.pdf} 
\caption{ (a) represents the terahertz spectra generated by 0.5mm (blue) and 4 mm OP (red) respectively. The spatial dependence of the terahertz spectral densities are shown in (b) and (c).\label{simu_nles_spec}}
\end{figure}
 \vspace{-0.5cm}
Figure \ref{simu_nles_eff} shows the dependence of the conversion efficiency, the terahertz fluence and the spectra on the OP fluence. Lower OP fluence leads to a slightly larger terahertz beam size. The terahertz spectrum practically does not depend on the fluence.  The OP output spectrum and fluence corresponding to Figs. 2(a) and 4(b2) are given in Fig. \ref{nles_op}.  

\begin{figure}[H]
\centering
\includegraphics[width=0.5\textwidth]{nles_energy_compare.pdf} 
\caption{With a given OP beam size ($\sigma_x'=$0.5 mm), (a) represents the internal terahertz conversion efficiency with pump fluence 70.7 mJ/cm$^2$, 35.4 mJ/cm$^2$ and 17.7 mJ/cm$^2$ respectively. The terahertz fluences and corresponding spectra at the peak of each efficiency curve are shown in (b) and (c) respectively. The out-coupling energy transmission is denoted by T. \label{simu_nles_eff}}
\end{figure}
\vspace{-0.5cm}
\begin{figure}[H]
\centering
\includegraphics[width=0.5\textwidth]{nles_op.pdf} 
\caption{The OP spectral density (a), spectrum (b) and the fluence (c) are presented.\label{nles_op}}
\end{figure}
 \vspace{-1.0cm}
\subsection*{(c) reflective echelon (RES)}
The echelon step sizes in $z'$ and $x'$ dimensions are represented by H, W respectively (see Fig. 1(c) in the main text). By setting  $\sigma_m' \sim \text{ W}/2$, one can write $q_0=i\pi\sigma_m'^2/\lambda_0$ and $q_m=q_0-2m\text{H}$. As a result, the OP electric field behind the LN input surface is shown in Eq. (\mbox{\ref{res_1}}), where $\sigma_x'=\sigma'f_2 /f_1$. The second order polarization can be written as Eq. (\mbox{\ref{res_2}}).
\begin{eqnarray}
E(\omega,x',z')=E_0\sum _{m=-\infty} ^{\infty}\sqrt{\frac{f_1 q_0}{f_2 q_m}}  \exp \ls \frac{-x'^2f_1^2}{2\sigma_x'^2f_2^2}  \rs &&  \nonumber \\
\times  \exp \lsq \frac{ -ik_0(x'+m\text{W}f_2/f_1)^2f_1^2}{2q_m f_2^2}\rsq  \exp \ls -\frac{\Delta \omega^2\tau^2}{4} \rs&& \nonumber \\
 \times \exp \ls i2\Delta \omega m\text{H}/c \rs\exp \lsq \frac{-i\omega n_{ir}(\omega)z'}{c} \rsq  && \label{res_1} 
 \end{eqnarray}
 \begin{eqnarray}
P(\Omega,x',z')=\frac{E_0^2\chi^{(2)}\sqrt{2\pi}f_1 }{\tau f_2 }\sum _{m=-\infty} ^{\infty}\abs{\frac{q_0}{q_m}}  \exp \lsq  \frac{-x'^2 f_1^2}{\sigma'^2 f_2^2  } \rsq  &&   \nonumber\\ 
 \times \exp{\ls \frac{-\Omega^2\tau^2}{8} \rs} \exp \lsq \frac{ -k_0^2\sigma_m'^2(x'+m\text{W}f_2/f_1)^2f_1^2}{\abs{q_m}^2 f_2^2}\rsq  &&   
 \nonumber \\
\times \exp{\lsq iz'\Omega (n_t\cos{(\gamma)}-n_g)/c \rsq}  && \nonumber \\
\times \exp{\lc i\Omega  \lsq n_t \sin{(\gamma)}x'+2m\text{H} \rsq /c\rc} \hspace{0.5cm}  \label{res_2}  &&
\end{eqnarray}
 {From the last two exponential terms in Eq. (\mbox{\ref{res_2}}), it can be seen that the PM conditions are }
\begin{eqnarray}
&&  n_t\cos{(\gamma)}=n_g, \nonumber\\
&&  \tan{(\gamma)}x'=-2m\text{H}/n_g.\label{sup_res_pm}
\end{eqnarray}
Before the imaging system, the delay between the beamlets in $z'$ dimension is $2\text{H}/c$ and $m=1$ represents the beamlet centered at $x'=$W. After the imaging system, the delay remains $2\text{H}/c$, whereas for $m=1$ the beamlet is centered at position $x'=-\text{W}f_2/f_1$. Consequently, together with Eq. (\ref{sup_res_pm}), it can be seen that the echelon structure needs to satisfy the condition, $\tan{(\gamma)}=2\text{H}f_1/(\text{W}n_gf_2)$ as indicated in the main text, to generate terahertz effectively. 

\begin{figure}[H]
\centering
\includegraphics[width=0.5\textwidth]{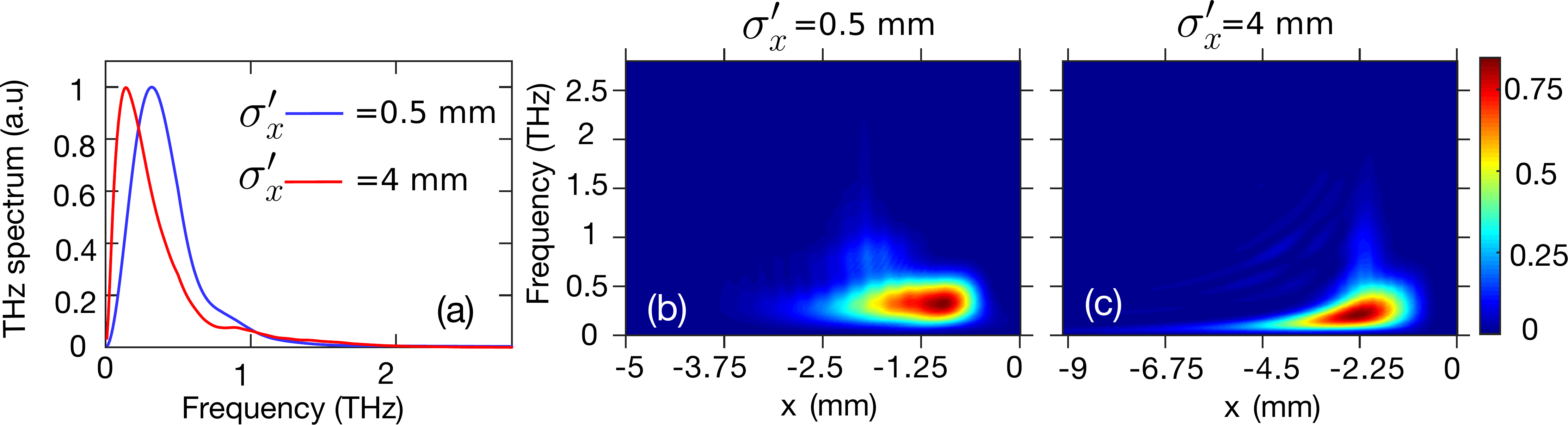} 
\caption{ (a) represents the terahertz spectra generated by 0.5 mm (blue) and 4 mm OP (red) respectively. The spatial dependence of the terahertz spectral densities are shown in (b) and (c).\label{simu_res_spec}}
\end{figure}

\begin{figure}[H]
\centering
\includegraphics[width=0.5\textwidth]{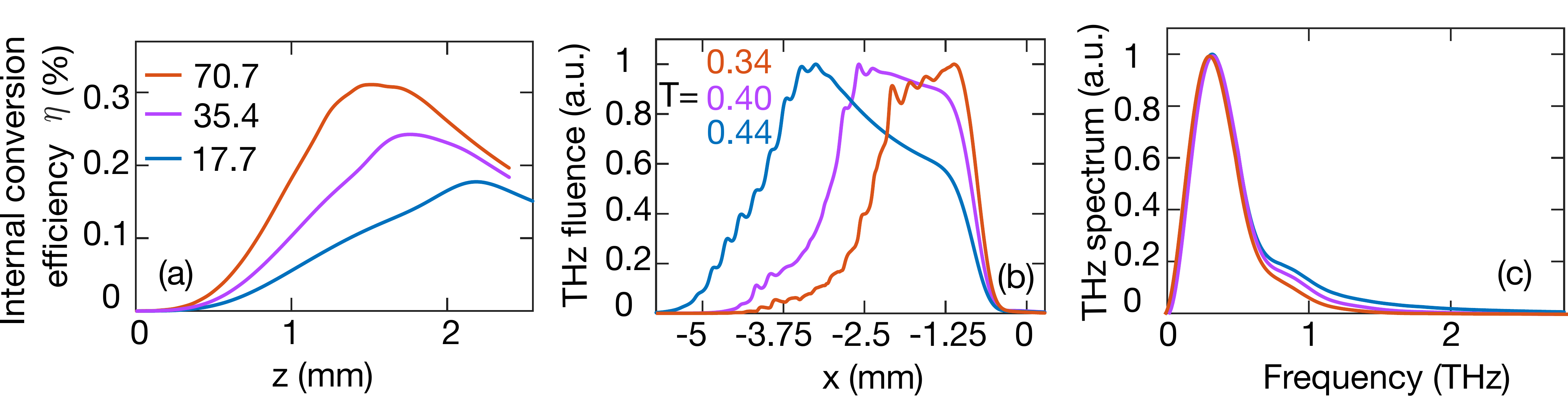} 
\caption{With a given OP beam size ($\sigma_x'=$0.5 mm), (a) represents the internal terahertz conversion efficiency with pump fluence 70.7 mJ/cm$^2$, 35.4 mJ/cm$^2$ and 17.7 mJ/cm$^2$ respectively. The terahertz fluence and corresponding spectra at the peak of each efficiency curve are shown in (b) and (c) respectively. The out-coupling energy transmission is denoted by T. \label{simu_res_eff}}
\end{figure}

The terahertz spectra generated by different OP beam sizes are presented in Fig. \ref{simu_res_spec}. The terahertz pulse generated by larger OP beam size possesses lower center frequency due to the absorption and longer interaction length. Figure \ref{simu_res_eff} shows the dependence of the conversion efficiency, the terahertz fluence and the spectra on the OP fluence. Lower OP fluence leads to larger terahertz beam size. The terahertz spectrum does not vary much with respect to the OP fluence. The OP output spectrum and fluence correspond to Figs. 2(a) and 4(c2) are given in Fig. \ref{res_op}.  

\vspace{-0.5cm}
\vspace{-0.5cm}
\begin{figure}[H]
\centering
\includegraphics[width=0.5\textwidth]{echelon_op.pdf} 
\caption{The OP spectral density (a), spectrum (b) and the fluence (c) are presented.\label{res_op}}
\end{figure}
 \vspace{-0.5cm}
\subsection*{(d) multi-step phase mask (MSPM)}
For this scheme, no imaging system is required and thus $\sigma'=\sigma_x'$. The mask step sizes in $z'$ and $x'$ dimensions are represented by H, W respectively (see Fig. 1(d) in the main text). The OP electric behind the LN surface is presented in Eq. (\ref{mspm_1}), where $\sigma_m' \sim \text{ W}/2$,  $q_0=i\pi\sigma_m'^2/\lambda_0$,   $q_m=q_0+L_0+L_1/n-m\text{H}(1-1/n)$ and $n$ is the refractiv index of the mask material. The term $L_0+L_1n$ is the optical path from the center of the OP ($x'=0$) to the LN surface, with the propagation distances in the air and in the MSPM noted by $L_0$ and $ L_1$ respectively. Consequently, the second order polarization of the terahertz generation is shown in Eq. (\ref{mspm_p}).
\begin{eqnarray}
E(\omega,x',z')=E_0\sum _{m=-\infty} ^{\infty}\sqrt{\frac{q_0}{q_m}} \exp \ls -\frac{\Delta \omega^2\tau^2}{4} \rs \exp \ls \frac{-x'^2}{2\sigma_x'^2}  \rs  && \nonumber \\
\times \exp \lsq \frac{ -i(x'-m\text{W})^2k_0}{2q_m }\rsq  \exp \ls \frac{-i\omega n_{ir}(\omega)z'}{c} \rs  && \nonumber \\
\times \exp \lc {-i\Delta \omega \lsq m \text{H} (n-1)+L_0+L_1n\rsq}/{c}   \rc &&\label{mspm_1}
\end{eqnarray}
\begin{eqnarray}
P(\Omega,x',z')=\frac{E_0^2\chi^{(2)}\sqrt{2\pi} }{\tau  }\sum _{m=-\infty} ^{\infty}\abs{\frac{q_0}{ {q_m}}}  \exp \ls \frac{-\Omega^2\tau^2}{8} \rs  &&  \nonumber  \\
\times \exp \lsq -i\Omega \ls L_0+L_1n\rs/c \rsq  \exp \ls \frac{-x'^2}{\sigma_x'^2} \rs && \nonumber \\
\times  \exp \lsq \frac{-k_0^2\sigma_m'^2(x'-m\text{W})^2}{\abs{q_m}^2}  \rsq  &&\nonumber \\
 \times \exp \lsq  i\Omega z'(n_t\cos{(\gamma)}-n_g)/c \rsq  && \nonumber \\
 \times \exp \lc i\Omega \lsq n_t\sin{(\gamma)}x'-m\text{H}(n-1)\rsq/c \rc &&  \label{mspm_p}
\end{eqnarray}
In the argument of the first exponential term in Eq. (\ref{mspm_p}), {$i\Omega \ls L_0+L_1n\rs/c$} represents a constant time delay and thus, can be set to zero without losing generality. From Eq. (\ref{mspm_p}), {one can obtain the PM conditions as below}.
\begin{eqnarray}
&& n_t\cos{(\gamma)}=n_g, \nonumber\\
&& n_g\tan{(\gamma)}x'=m\text{H}(n-1) \label{sup_mspm_pm}
\end{eqnarray}
When $m=1$, i.e $x'=$W, Eq. (\ref{sup_mspm_pm}) reduces to $n_g\tan{(\gamma)}/(n-1)=\text{H}/\text{W}$,{ which is given in the main text. }

The terahertz spectra generated by different OP beam sizes are presented in Fig. \ref{simu_mspm_spec}. The terahertz spectrum generated by large OP beam size possesses a smooth distribution whereas the terahertz spectrum generated by small OP beam size indicates an interference phenomenon. Figure \ref{simu_silicon_eff} shows the conversion efficiency, the terahertz fluence and the spectra on the OP fluence. Lower OP fluence leads to larger terahertz beam size. The terahertz spectrum is independent of the OP fluence.  The OP output spectrum and fluence correspond to Figs. 2(a) and 4(d2) are given in Fig. \ref{silicon_op}.  

\begin{figure}[H]
\centering
\includegraphics[width=0.5\textwidth]{silica_spec.pdf} 
\caption{ (a) represents the terahertz spectra generated by 0.5 mm (blue) and 4 mm OP (red) respectively. The spatial dependence of the terahertz spectral densities are shown in (b) and (c) respectively.\label{simu_mspm_spec}}
\end{figure}
\vspace{-0.5cm}
\begin{figure}[H]
\centering
\includegraphics[width=0.5\textwidth]{silicon_energy_compare.pdf} 
\caption{With a given OP beam size ($\sigma_x'=$0.5 mm), (a) represents the internal terahertz conversion efficiency with pump fluence 70.7 mJ/cm$^2$, 35.4 mJ/cm$^2$ and 17.7 mJ/cm$^2$ respectively. The terahertz fluences and corresponding spectra at the peak of each efficiency curve are shown in (b) and (c) respectively. The out-coupling energy transmission is denoted by T. \label{simu_silicon_eff}}
\end{figure}
\vspace{-0.5cm}
\begin{figure}[H]
\centering
\includegraphics[width=0.5\textwidth]{silicon_op.pdf} 
\caption{The OP spectral density (a), spectrum (b) and the fluence (c) are presented. \label{silicon_op}}
\end{figure}
 \vspace{-0.5cm}
\subsection*{(e) spatial temporal chirp (STC)}
In this scheme, $\sigma'=\sigma_x'$. The input electric field and the second order nonlinear polarisation are shown in Eqs. \ref{sup_stc1} and \ref{sup_stc_pol}, respectively.
\begin{eqnarray}
 E(\omega,x',z')=E_0\exp\left(\frac{-x'^2}{2\sigma_x'^2}\right)\exp \left[-\frac{(\Delta \omega-vx')^2\tau^2}{4}\right]&& \nonumber \\
 \times \exp\left(-\frac{i\phi_2\Delta \omega^2}{2}-ik(\omega)z'\right)\label{sup_stc1} && 
 \end{eqnarray}
 \begin{eqnarray}
   P(\Omega,x',z')=\chi^{(2)}\sqrt{2\pi}\frac{E_0^2}{\tau} \exp\left(-\frac{x'^2}{\sigma_x'^2}\right)&& \nonumber\\
   \times \exp \left[-\frac{\Omega^2\tau^2}{8}\left(1+\frac{4\phi_2^2}{\tau^4}\right)\right]&& \nonumber \\
 \times \exp\left\{
 \frac{-i\Omega z'}{c}\left[n_g-n_t\cos(\gamma)\right]\right\} &&\nonumber \\
  \times \exp{\left\{-i\Omega x'\left[v\phi_2-\frac{n_t\sin(\gamma)}{c}\right]\right\}} && \label{sup_stc_pol}
\end{eqnarray}

The PM conditions are 
\begin{eqnarray}
 && n_t\cos{(\gamma)}=n_g, \nonumber\\
&& v\phi_2=\tan{(\gamma)}n_g/c.\label{stc_pm}
\end{eqnarray}

\begin{figure}[H]
\centering
\includegraphics[width=0.35\textwidth]{st_spec.pdf} 
\caption{(a) represents the terahertz spectrum generated by 0.5 mm OP. The spatial dependence of the terahertz spectral density is shown in (b).\label{simu_st_spec}}
\end{figure}
\vspace{-0.5cm}
The terahertz spectrum and spectral density generated by OP beam size $\sigma_x'=0.5$ mm are presented in Fig. \ref{simu_st_spec}. Figure \ref{simu_st_eff} shows the dependence of the conversion efficiency, the terahertz fluence and the spectra on the OP fluence. Lower OP fluence leads to larger terahertz beam size. The terahertz spectrum is independent of the OP fluence. The OP output spectrum and fluence correspond to Figs. 2(a) and 4(e2) are given in Fig. \ref{st_op}.
\begin{figure}[H]
\centering
\includegraphics[width=0.5\textwidth]{st_energy_compare.pdf} 
\caption{With a given OP beam size ($\sigma_x'=$0.5 mm), (a) represents the internal terahertz conversion efficiency with pump fluence 70.7 mJ/cm$^2$, 35.4 mJ/cm$^2$ and 17.7 mJ/cm$^2$ respectively. The terahertz fluences and corresponding spectra at the peak of each efficiency curve are shown in (b) and (c) respectively. The out-coupling energy transmission is denoted by T. \label{simu_st_eff}}
\end{figure}
\vspace{-0.5cm}
\begin{figure}[H]
\centering
\includegraphics[width=0.5\textwidth]{st_op.pdf} 
\caption{The OP spectral density (a), spectrum (b) and the fluence (c) are presented.\label{st_op}}
\end{figure}

\bibliographystyle{lpr}
\bibliography{sample}